\documentstyle[prb,aps,eqsecnum]{revtex}
\begin{document}
\tighten
\input{psfig}
\psfull

\title{A study of energy minimization
techniques applied to protein design}
\author{Cristian Micheletti and Amos Maritan}
\address{International School for Advanced Studies (S.I.S.S.A.),
Via Beirut 2-4, 34014 Trieste, Italy, \\
Istituto Nazionale per la Fisica della
    Materia\\
and The Abdus Salam International Centre for Theoretical Physics} 
\date{\today}
\maketitle
\begin{abstract}
We present a detailed study of the performance and reliability of
design procedures based on energy minimization. The analysis is
carried out for model proteins where exact results can be obtained
through exhaustive enumeration. The efficiency of design techniques is
assessed as a function of protein lengths and number of classes into
which amino acids are coarse grained. It turns out that, while energy
minimization strategies can identify correct solutions in most
circumstances, it may be impossible for numerical implementations of
design algorithms to meet the efficiency required to yield correct
solutions in realistic contexts. We also investigated how the design
efficiency varies when putative solutions are required to obey some
external constraints and found that a restriction of the sequence space
impairs the design performance rather than boosting it. Finally some
alternate design strategies based on a correct treatment of the free
energy are discussed. These are shown to be significantly more
efficient than energy-based methods while requiring nearly the same
CPU time.
\end{abstract}

\section{Introduction}

A few decades ago C. Anfinsen [\onlinecite{a}] showed that the structural
information of naturally occurring proteins is entirely encoded by the
corresponding amino acid sequence. Since then many biologists,
chemists and physicists have spent their efforts in trying to identify
and simulate the mechanisms through which a given sequence reaches its
stable, native conformation (protein folding)
[\onlinecite{b0,b1,b2}]. The inverse problem, also known as protein
design, has similarly resisted the efforts of an ever-growing number
of researchers who are now tackling it with an arsenal of techniques
including {\em ab initio} molecular dynamics and concepts of
theoretical physics
[\onlinecite{i14,i10,i2,i4,M98b,i12,i15,i13,i3,i11,S98,i8}].  The
complexity of the problem is enormous because, in principle, it
entails an exhaustive comparison of the native states of all sequences
in search for the one(s) matching the desired target structure
[\onlinecite{i10,i12,i11}].

This problem has been recently formulated into a general mathematical
form appropriate for numerical implementation [\onlinecite{i11}] which
shows that solving the design problem for a structure, $\Gamma$,
amounts to the identification of the amino acid sequence, $S$, that
maximizes the occupation probability, $P_\Gamma(S)$,

\begin{equation}
P_\Gamma(S) \equiv { e^{- \beta E_\Gamma(S)} \over
\sum_{\Gamma^\prime} e^{- \beta E_{\Gamma^\prime}(S)}} =
  e^{- \beta \{ E_\Gamma(S) - F(S) \} }
\label{eqn:occprob}
\end{equation}

\noindent where $\beta$ is the Boltzmann factor, $E_\Gamma(S)$ is the
energy of the sequence $S$ over the structure $\Gamma$ and the sum in
the denominator is taken over all possible structures, $\{
\Gamma^\prime \}$ having the same length of $\Gamma$. The winning
sequence will maximize $P_\Gamma(S)$ at all temperatures below the
folding transition temperature (where the occupation probability of
the native state is macroscopic). The occupation probability
$P_\Gamma(S)$ has the following simple physical interpretation. The
quantity $F(S)$ implicitly defined in (\ref{eqn:occprob}) corresponds
to the free energy of sequence $S$. Below the folding transition the
dominant contribution to $F(S)$ comes from the ground state(s) of
$S$. Hence, maximising the functional $P_\Gamma(S)$ at low temperature
is equivalent to identifying the sequences for which $E_\Gamma(S)$
corresponds to the ground state energy and the ground state degeneracy
is the lowest possible.

Several obstacles need to be overcome to implement
Eq. (\ref{eqn:occprob}). First it is necessary to know the form of
${\cal H}$, i.e. the amino acid interaction potentials. Second the
calculation of $F(S)$ for a given sequence entails, in principle, a
complete exploration of the conformation space. Finally, the
quantities $E_\Gamma(S)$ need to be calculated for all
sequences $S$ in order to find the ones maximising
Eq. (\ref{eqn:occprob}).

The exploration of the sequence space is rather easy to carry out and
may simply involve a generation of random sequences; furthermore the
dimension of the sequence space is often restricted by grouping the 20
different amino acids occurring in nature into a reduced number of
classes according to their chemical similarities. Instead, the sum
over alternative conformations, $\{\Gamma^\prime\}$ in
(\ref{eqn:occprob}) requires the generation of physically stable
structures that compete significantly with the true native state of
$S$ to be occupied at low temperature. These give the most significant
contribution to $F(S)$ below the folding transition temperature.

This problem is usually circumvented by neglecting the free
energy contribution in Eq. (\ref{eqn:occprob})
[\onlinecite{i8,new,SGZ,i9}]. Maximising $P_\Gamma(S)$ then corresponds to
identifying the sequence with lowest possible energy on $\Gamma$. This
procedure is, in principle, not guaranteed to yield the correct
answer. In fact, it may well be that the sequence having the smallest
energy on $\Gamma$ has an even lower energy on a different structure.

Despite the fact that the method is not rigorous it has encountered
some favour due to its simplicity of use. The aim of the present paper
is to investigate how efficiency/reliability of this procedure varies
as

\begin{enumerate}
\item{the number of classes into which amino acids are subdivided is
increased while the peptide length in held fixed,}
\item{the length is changed while the number of classes is fixed,}
\item{the amino acids in some positions are kept quenched while the others
are chosen to minimize the total energy.}
\end{enumerate}

These questions will be formulated in lattice contexts where an
impartial and rigorous assessment of design techniques can be found
with the aid of a computer. In particular we will recourse to
exhaustive enumeration whenever computational resources will allow
it. We will limit our structural ensemble to compact structures on a
square lattice. For lengths for which exact enumeration is feasible
(up to a few dozens of residues) the square lattice can yield a ratio
of exposed/buried residues much closer to the real case than the cubic
counterpart.

From the analysis presented here it will appear that, design
techniques based on energy minimization encounter growing limitations
as the number of amino acid classes and peptide length increase to
approach realistic values. We will also examine approximations to the
free energy that turn out to be more efficient than energy
minimization procedures and take up the same amount of computational
time.

Throughout this study we will adopt the following Hamiltonian
 
\begin{equation}
E_\Gamma(S) = \sum_{i,j} \Delta_{ij} \epsilon(\sigma_i,\sigma_j)
\end{equation}

\noindent where $S_i$ denotes the residue at position $i$, $\epsilon$
is the interaction matrix and $\Delta_{ij}$ is equal to 1 if $i$ and
$j$ are neighbouring residues that are not consecutive along the chain
and zero otherwise.

\section{Design by energy minimization}

Design techniques by energy minimization were first introduced by
E. I. Shakhnovich in the context of lattice models [\onlinecite{i8}].
This procedure has been justified within the approximation of the
discrete random energy model [\onlinecite{D81}] in reference
[\onlinecite{GS}].  One of the most stringent tests of this method was
carried out in a competition between the research teams of Harvard and
San Francisco [\onlinecite{i6}]. The goal was to design 10
three-dimensional compact structures of 48 beads within the HP
framework. The HP model consists of only two classes of amino acids,
hydrophobic [H] and polar [P]. A favourable contact energy,
$\epsilon_{HH}=-1$ was assigned to two non-consecutive residues which
are one lattice spacing apart, while the other interactions,
$\epsilon_{PH}$, $\epsilon_{HP}$, $\epsilon_{PP}$ were set equal to
zero. These values are defined modulus a sufficiently negative
additive constant to guarantee that the ground states are all compact.
This model favours the collapse of a hydrophobic core and is thought
to mimic the main driving force of protein folding
[\onlinecite{i1,i5,i15a}].

The design strategy followed by the Harvard group was to consider only
sequences with the same number of H and P residues (equal composition)
and then identify, among them, the sequences having minimum energy on
each target structure. Disappointingly a correct answer was found in
only one of ten cases [\onlinecite{i6}].

Without the restriction that solutions have equal number of H and P
residues, $n_H=n_P=24$, the energy minimization procedure would have
yielded sequences with large values of $n_H$ (so that a couple of $H$
residues was present in correspondence of each geometrical contact of
the target structure).  These solutions, like their counterpart where
all residues are assigned as polar, correspond to trivial answers to
the design problem since they have an enormous ground state
degeneracy.

A correct solution to all 10 Harvard-San Francisco problems was found
recently by a careful treatment of the free energy in
(\ref{eqn:occprob}) [\onlinecite{i18}]. The study also showed that, to a
good extent, the free energy of sequences with the same composition is
approximately constant. In this case, for a given composition,
maximising (\ref{eqn:occprob}) corresponds to minimising the
energy. Therefore, provided that solutions exist at a given
composition, energy minimization techniques may be apt to find
them. One question that arises naturally is: which fraction of
solutions having a given concentration can be found using energy
minimization techniques?

\subsection{Two classes of amino acids.}

We will answer this and other questions by considering first the case
where the amino acids are subdivided into two classes. Two different
interaction matrices will be used in order to identify the qualitative
features of the energy minimization procedures that do not depend on
the details of the model. The first choice of parameters corresponds
to the standard HP model [\onlinecite{i15a}] while, for the second one, an
interaction matrix previously adopted by the NEC group will be
used [\onlinecite{Li}].

\begin{figure}
\centerline{\psfig{figure=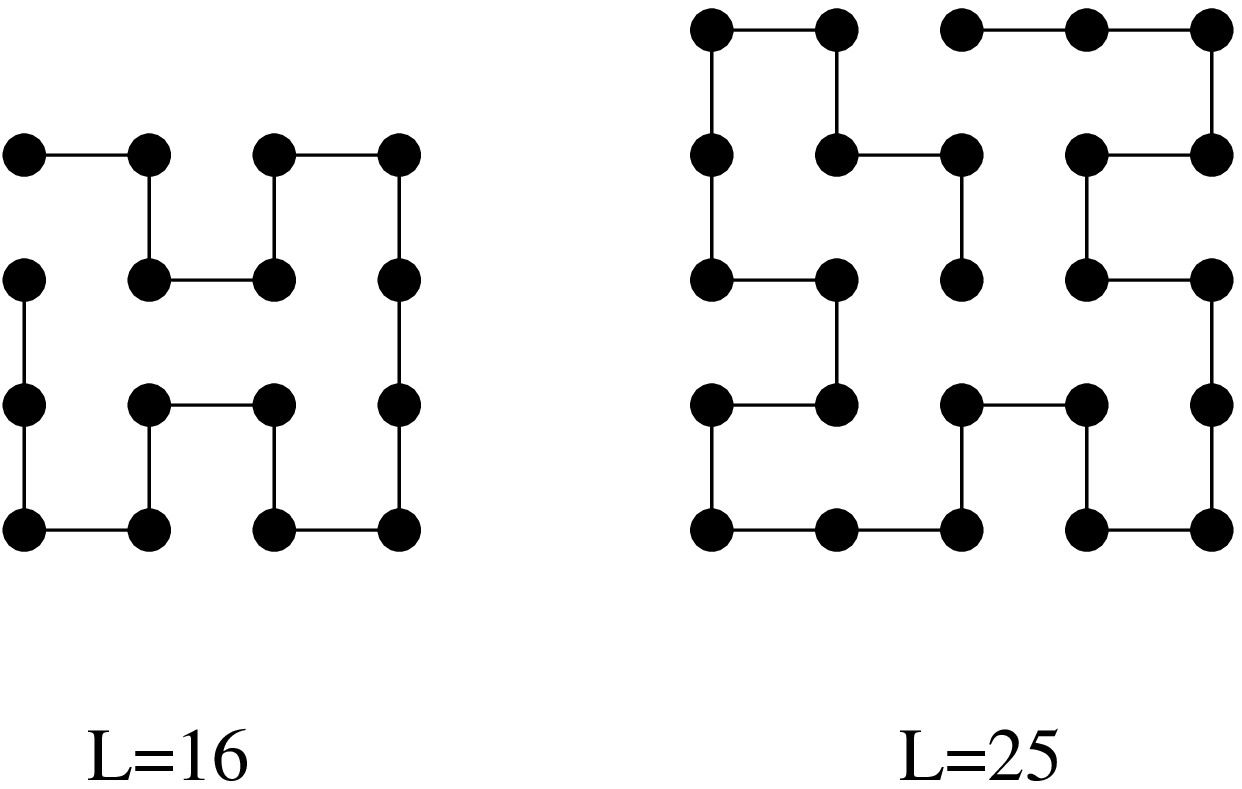,width=3.0in}}
\caption{The most encodable compact structures among chains of length
a) 16, b) 25.}
\label{fig:encod}
\end{figure}

In order to collect good statistics it was decided to perform the
design study on the most encodable compact structures, i.e. compact
structures that are designed by the highest number of sequences
[\onlinecite{Li}]. Such structures have been shown to display a high degree
of geometrical regularity mimicking that found in real proteins. As a
by-product of our analysis we found that the encodability property is
robust against changes of the model like energy interactions or number
of amino acid classes and confirm that the property of encodability
has mainly a geometrical origin [\onlinecite{Li,M98c}]. For example we found
that the most designable structures of length 16 and 25 (see
Fig. \ref{fig:encod}) remained the same when using the the HP
parameters or the NEC ones.

The main tool used for the analysis was a double backtracking
algorithm, which mounted every sequence of length $L=16$ with $n_H=8$
(they are nearly 13000) on each of the 69 compact
conformations. Working at fixed concentration, we then calculated the
number, $n_\Gamma(E)$, of sequences that admitted $\Gamma$ as their
unique ground state and have energy $E$, as well as the total number
of sequences,$N_\Gamma(E)$, which attain an energy $E$ when mounted on
$\Gamma$, irrespective of what their ground state is.

\begin{figure}
\centerline{\psfig{figure=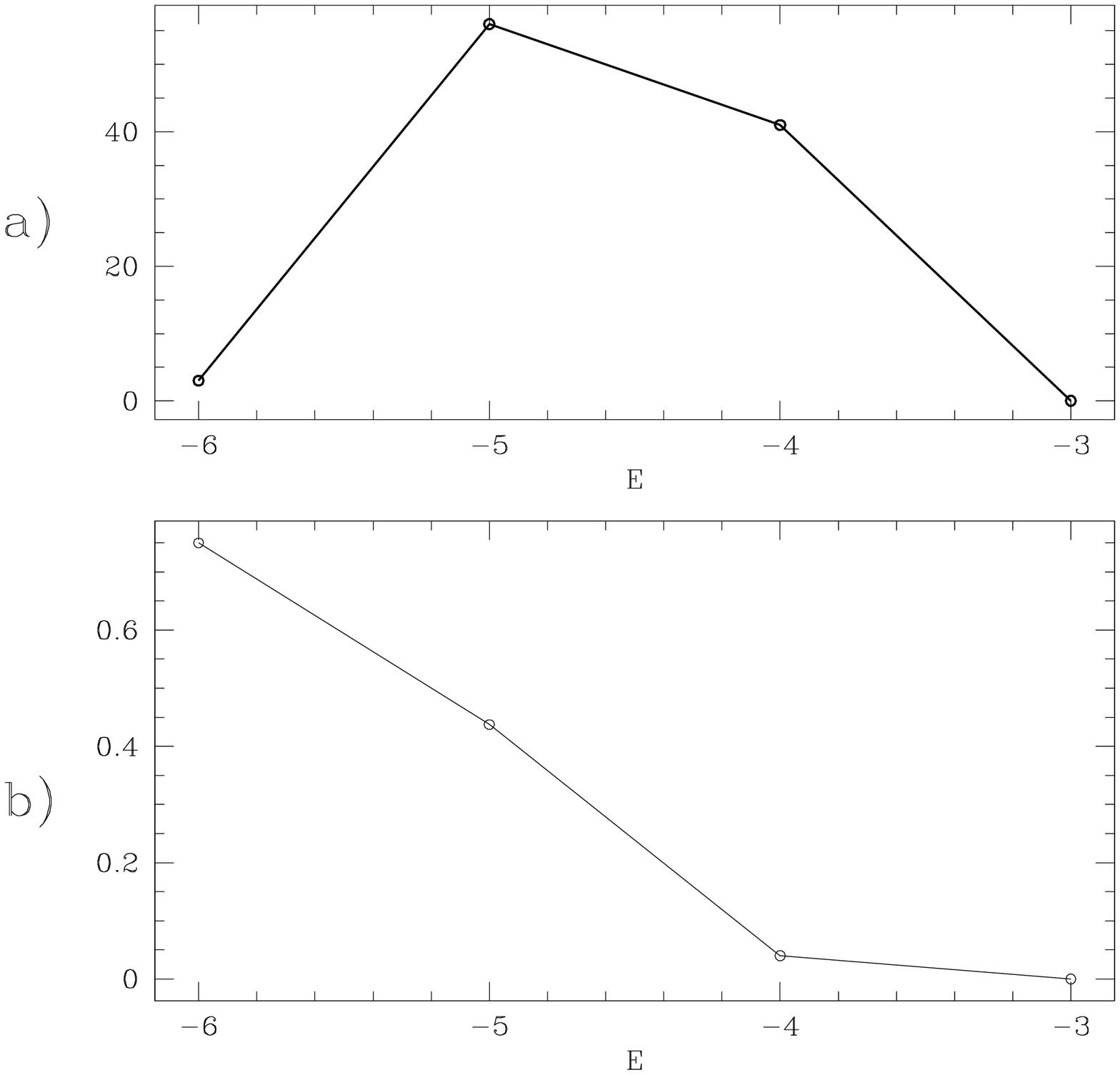,width=3.0in}}
\caption{Distribution of a) $n_\Gamma(E)$, b)
$n_\Gamma(E)/N_\Gamma(E)$ for sequences of length 16
and $n_H=8$ with HP-type interactions. The target structure $\Gamma$
is shown in Fig. \protect{\ref{fig:encod}}a).}
\label{fig:hp16}
\end{figure}

The behaviour of $n_\Gamma(E)$ is shown in the upper plot of Fig.
\ref{fig:hp16}, while in Fig. \ref{fig:hp16}b) is sketched the ratio
$n_\Gamma(E)/N_\Gamma(E)$. Interestingly $n_\Gamma(E)$ has the shape
of a bell and shows that only a small fraction of solutions, 3 out of
100, have minimum energy $E=-6$ since the overall number of sequences
having minimum energy is four so the fraction of them which are a
correct solution to the problem is 0.75.

\begin{figure}
\centerline{\psfig{figure=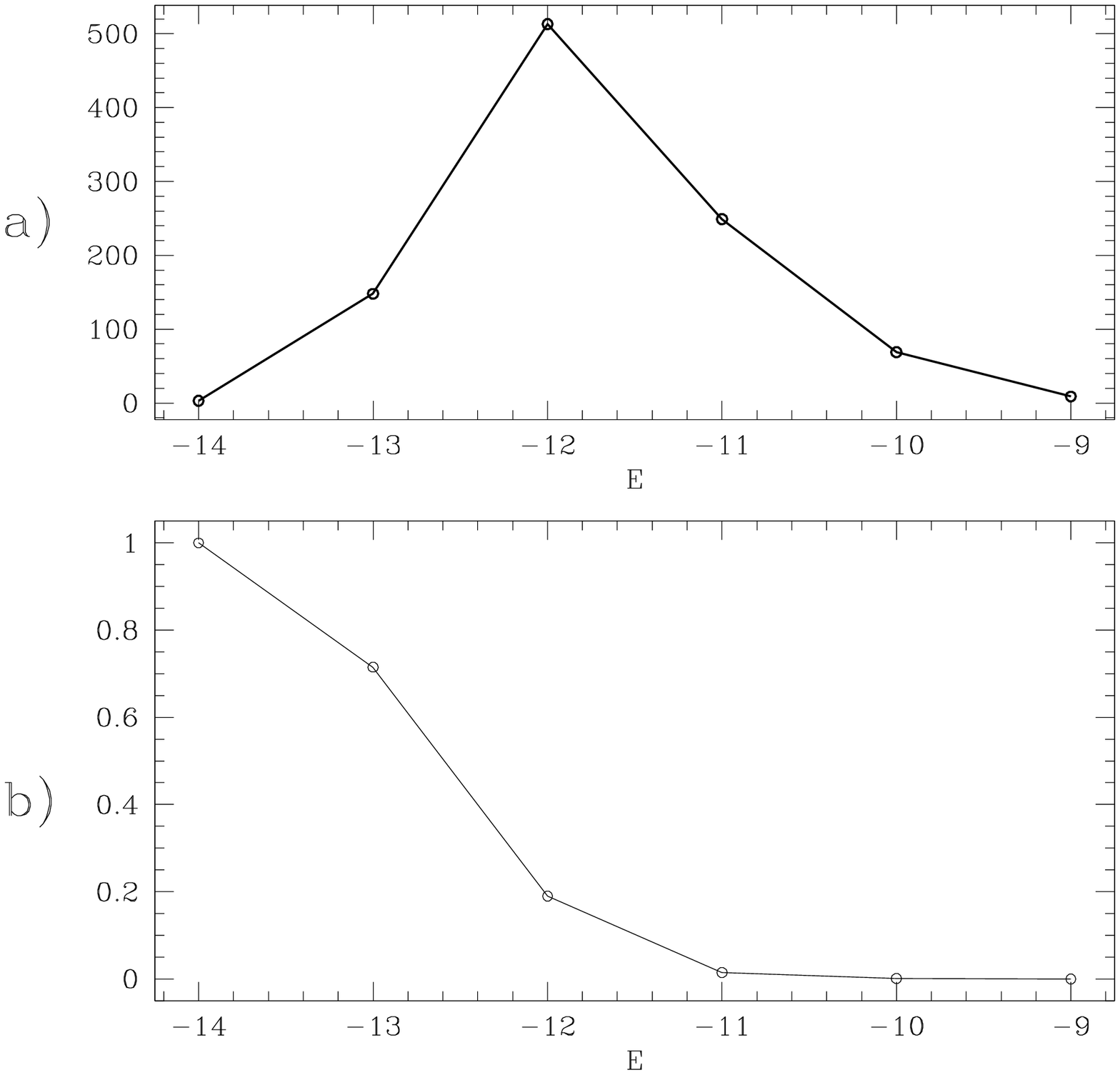,width=3.0in}}
\caption{Distribution of a) $n_\Gamma(E)$, b)
$n_\Gamma(E)/N_\Gamma(E)$ for sequences of length 25
and $n_H=16$ with HP-type interactions. The target structure $\Gamma$
is shown in Fig. \protect{\ref{fig:encod}}b) }
\label{fig:hp25}
\end{figure}

\begin{figure}
\centerline{\psfig{figure=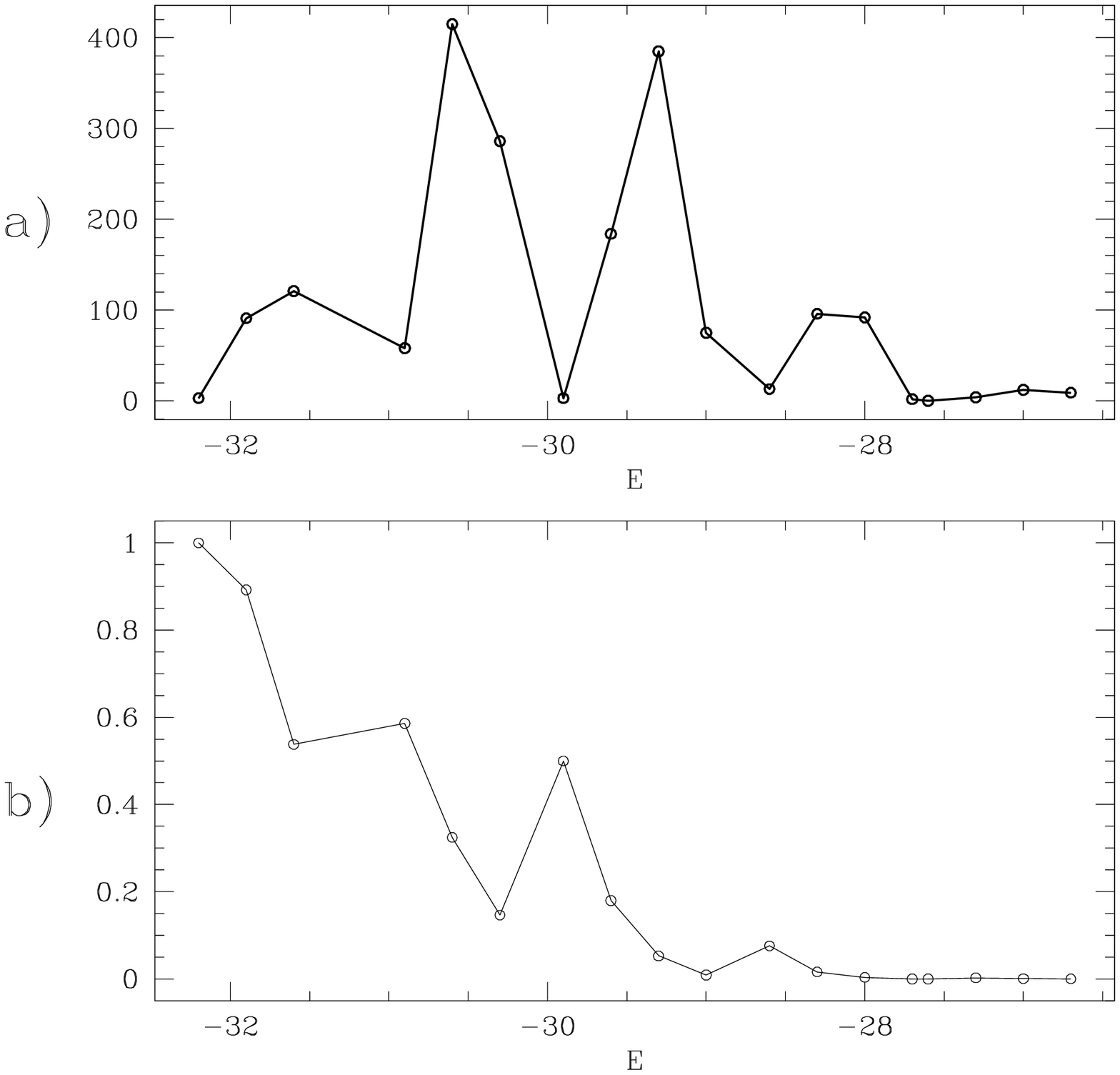,width=3.0in}}
\caption{Distribution of a) $n_\Gamma(E)$, b)
$n_\Gamma(E)/N_\Gamma(E)$ for sequences of length 25
and $n_H=16$ with NEC-type interactions. The target structure $\Gamma$
is shown in Fig. \protect{\ref{fig:encod}}b) }
\label{fig:nec25}
\end{figure}

In Fig. \ref{fig:hp25} we have represented analogous results for the
$L=25$ case. The sequences were constrained to have $n_H=16$; there
are approximately $10^6$ such sequences, while the number of compact
structures is 1081. Analogous enumerations were carried out for the
same sequence lengths and compositions but adopting the NEC
interaction parameters, $\epsilon_{HH}=-2.3$,
$\epsilon_{PH}=\epsilon_{HP}=-1.0$, $\epsilon_{PP}=0$. The results
were qualitatively similar to those of the pure HP model, a part from
the fact that the overall number of design solutions was greatly
enhanced. For these reasons we present only the results for length 25
and $n_H= 16$, as shown in Fig. \ref{fig:nec25}. It can be seen that
$n_\Gamma(E)$ presents an oscillatory behaviour. This is due to the
fact that there is a small number of amino acid classes and that the
entries of the interaction matrix have similar strength. In fact, two
closely spaced energy levels could be obtained using very different
sets of contact pairs; the ocillations of $n_\Gamma(E)$ reflect the
fact that the number of sequences contributing to the two sets of
contacts may vary significantly.

\subsection{Three and four classes of amino acids.}

Finally the case where the amino acids are subdivided in 3 and 4
classes was addressed. It is often remarked that the shortcomings of
energy minimization routines observed in HP-like contexts are due to
the artificially large ground state degeneracy
[\onlinecite{i6}]. Introducing more classes of amino acids will
tipically remove this artifact and may possibly lead to an improved
performance of energy-minimization schemes.

Due to the large increase of sequence-space volume an exhaustive
enumeration is feasible only for chains of length 16. Our interaction
matrix for the 3 classes case was the following,

\begin{equation}
\left(
\begin{array}{ccc}
 0   & -0.5 & -1.2 \\
-0.5 & -1.15 & -1.7 \\
-1.2 & -1.7 & -2.6 \\
\end{array}
\right)\ .
\label{eqn:3c}
\end{equation}

\noindent The entries in (\ref{eqn:3c}) were chosen so that the
segregation principle is satisfied. For symmetric interactions this
corresponds to requiring that

\begin{equation}
\epsilon_{ii} +\epsilon_{jj} \le 2 \epsilon_{ij},
\end{equation}

\noindent a property that is satisfied to a large extent by extracted
potentials like Miyazawa-Jernigan [\onlinecite{MJ}] or Maiorov-Crippen
[\onlinecite{MC}]. The matrix (\ref{eqn:3c}) may be regarded as an
extension of the NEC one, since the submatrice corresponding to the
interaction between the first two types of residues is equal, a part
from a scaling factor of $1/2$, to that used in reference
[\onlinecite{Li}].

The requirement to use a fixed concentration entails the subdivision
of sequences into 153 distinct bins. The performance of design
algorithms based on energy minimization is not uniform across bins; in
particular, for some concentrations, the method may fail to find
solutions. For two classes of amino acids this occurs, for example,
for $L=16$, $n_H=9$ and NEC parameters, where no correct solution can
be identified with the energy minimization despite the existence of
190 solutions with that length and composition. The optimal bin for
the analysis were chosen so that they contained the highest number of
solutions. This insures the collection of the best possible statistics
on the behaviour of $n_\Gamma(E)$ and $n_\Gamma(E)/N_\Gamma(E)$. For
three classes, one of the most populated bins corresponded to nearly
equal composition: $n_1 = 5,\ n_2= 5,\
n_3 = 6$, where $n_i$ denotes the number of residues of type $i$. The
results are shown in Fig. \ref{fig:3c16}.

Finally, for the 4 types case we extended the matrix (\ref{eqn:3c}) to

\begin{equation}
\left(
\begin{array}{cccc}
 0   & -0.5 & -1.2 & -1.3\\
-0.5 & -1.15 & -1.7 & -2.0\\
-1.2 & -1.7 & -2.6 & -2.7\\
-1.3 & -2.0 & -2.7 & -3.0\\
\end{array}
\right)\ '
\end{equation}

\noindent and chose to work with sequences with $n_1 = 7,\ n_2= 6,\
n_3 = 2,\ n_4=1$. At this composition there exist nearly $6 \cdot
10^5$ solutions; the behaviour of $n_\Gamma(E)$ and $n_\Gamma(E)/
N_\Gamma(E)$ is represented in Fig. \ref{fig:4c16}.

\section{Limitations of the energy minimization procedure}

The results presented so far show that energy minimization procedures
can be effective in selecting correct ground states of model proteins
of reasonable length and number of amino acids. This may come as a
surprise since, in principle, the method is not guaranteed to
work. Part of successes observed here were undoubtedly due to the
choice of highly encodable target structures.  Given that there exist
many sequences that are solution to the design problem (hundreds to
thousands according to length, number of classes etc.) it is plausible
that a handful of them will have very low values of energy. These will
be the (only) ones selected by energy minimization procedures. This
interpretation is corroborated by the fact that, when dealing with
target structures that are poorly encodable; (e.g. structures with 20 or
fewer design solutions), no correct answer to design can be normally
found through energy minimization schemes. This fact also sheds some
light on the failure of the Harvard attempts to solve the Harvard-San
Francisco problems. In fact, the target structures used on that
occasion were chosen at random and not according to designability
criteria. As argued in the original solution to the problem this was
also the reason why only intermediate (degenerate) solutions could be
found to all 10 Harvard-San Francisco problems.

In the rest of this section we will comment on the limitations that
affect energy minimization schemes even when they are adopted in very
favourable circumstances such as on designable compact structures.

The first limitation regards how much the curve
$n_\Gamma(E)/N_\Gamma(E)$ is ``squeezed'' against the minimum energy
boundary. In optimization procedures applied to realistic off-lattice
contexts where the number of classes and peptide lengths is too large
to allow a thorough search of the whole sequence space, the
minimization procedure will tipically come close to the lowest
possible energy but without reaching it. It is then paramount to
examine ``how close'' it is necessary to get to the ground state
energy to ensure that a significant fraction of the sequences having
that energy is a solution to the problem. As a quantitative measure we
introduce the parameter
\begin{equation}
x={\tilde{E} - E_{mim} \over E_{max} - E_{mim}}
\end{equation}

\noindent where $E_{min}$ and $E_{max}$ are respectively the minimum
and maximum energies for which solutions to the design problem exists
and $\tilde{E}$ is the energy below which a randomly picked sequences
has more than $50\%$ chance to be a solution to the design problem.
Thus, the lower the value of $x$  ($0 < x <1$) the worst is the
performance of the method.

Since the curves $n_\Gamma(E)/N_\Gamma(E)$ typically do not show a
smooth behaviour, $\tilde{E}$ is determined with the aid of a
high-order polynomial function interpolating $n_\Gamma(E)/N_\Gamma(E)$.
For the results of Fig.  \ref{fig:hp16} we have $x_{16}=0.38$.

Upon increasing the chain length the shoulder of the curve is shifted
closer to the minimum energy edge; in fact, from Fig. \ref{fig:hp25}
we have $x_{25}=0.28$. Finally, we considered the most encodable
structure of length 36 and considered sequences with the same length
and $n_H=18$. Since it is not feasible to mount all the sequences with
this composition on each of the 57337 compact structures we
resorted to a random sampling of $10^{7}$ sequences.  We obtained
$x_{36}=0.20$. The values of $x_{16},\ x_{25}$ and $x_{36}$ were
found not to change appreciably when using a different composition
provided that there exist a significant number of solutions.

\begin{figure}
\centerline{\psfig{figure=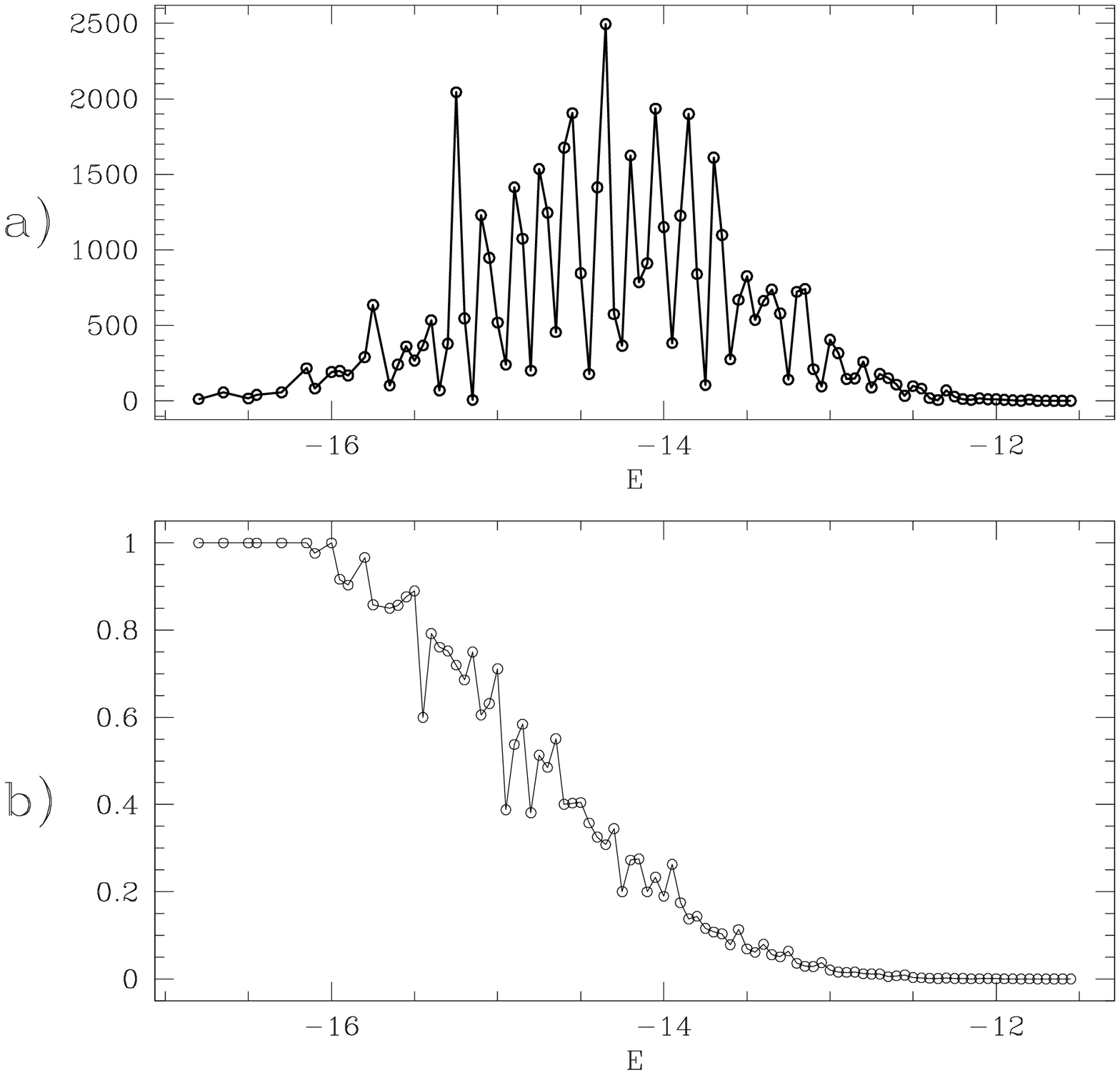,width=3.0in}}
\caption{Distribution of a) $n_\Gamma(E)$, b)
$n_\Gamma(E)/N_\Gamma(E)$ for sequences of length 16 and 3 classes of
amino acids interacting through (\protect{\ref{eqn:3c}}).  The target
structure $\Gamma$ is shown in Fig. \protect{\ref{fig:encod}}a).}
\label{fig:3c16}
\end{figure}

\begin{figure}
\centerline{\psfig{figure=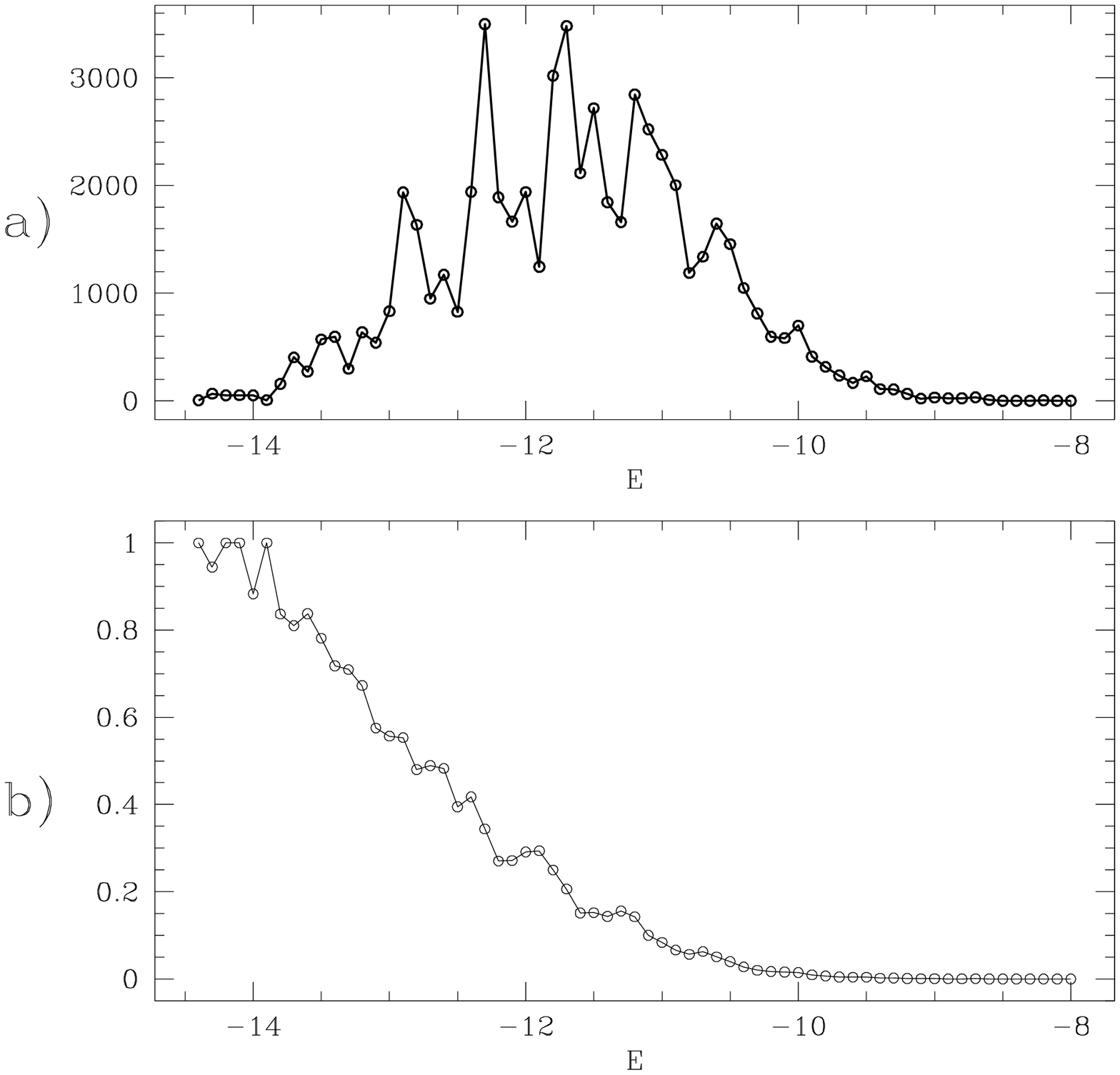,width=3.0in}}
\caption{Distribution of a) $n_\Gamma(E)$, b)
$n_\Gamma(E)/N_\Gamma(E)$ for sequences of length 16 and 3 classes of
amino acids interacting through (\protect{\ref{eqn:3c}}).  The target
structure $\Gamma$ is shown in Fig. \protect{\ref{fig:encod}}a).}
\label{fig:4c16}
\end{figure}

A similar trend can be observed by increasing the number of amino acid
classes. For the results of Figs. \ref{fig:3c16} and \ref{fig:4c16}
one has $x_{3c}=0.33$ and $x_{4c}=0.24$ showing a steady decrease as a
function of the number of classes (also remember that for two classes
we had $x_{16} = 0.4$) .

This shows that the demand on computational efficiency grows rapidly
as a function of length and classes. For realistic design on proteins
with a few hundred residues and 20 types of amino acids the required
efficiency may fall beyond the reach of computational techniques.

Nevertheless, even if it were possible to find the sequence(s) with
minimum energy, other issues need to be addressed.  In particular, a
limitation having far reaching consequences is that the number of
correct solutions that can be identified is only a tiny fraction of
the existing ones.  For example, for $L=16$, $n_H=8$ the fraction is
3/100 when using HP parameters and 3/207 for NEC ones. These
figures drop respectively to 24/1971 and 21/3978 for length 25,
$n_H=16$. The proportion decreases more dramatically when considering
more than two classes of amino acids, as can be seen in the plots of
Fig. \ref{fig:3c16} and Fig. \ref{fig:4c16}. In designing realistic
off-lattice proteins this feature is likely to pose severe limitations
to the reliability of the method.

In fact, it is expected that residues in naturally occurring sequences
were not selected on mere energetic considerations but also on
structural and biological functionality. These solutions would be
correctly identified when maximising the low temperature occupation
probability (\ref{eqn:occprob}) which, as said before, has 100 \%
efficiency throughout the energy range. On the contrary they could be
missed easily by energy-minimization schemes since sequences with
minimum energy on a given protein backbone may not be the most
suitable ones as far as biological function is concerned.

Yet energy minimization approaches could, in principle, still be
salvaged by arguing that the active sites of a protein are a small
fraction of the total residues and, when known, may be fixed {\em a
priori}. In an attempt to ``improve on nature'' (e.g. to increase the
thermodynamic stability of the native fold) the rest of the residue
could be found subsequently by energy minimization.

\begin{figure}
\centerline{\psfig{figure=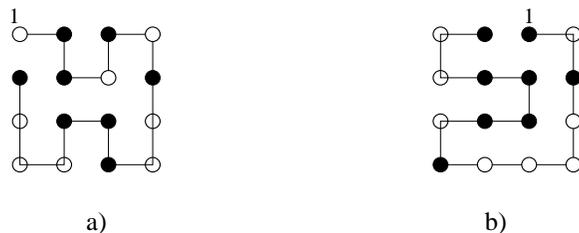,width=3.0in}}
\caption{a) a solution to the design problem on structure
\protect{\ref{fig:encod}} having energy $E=-13.2$. By quenching residues
4 and 14 and minimizing the energy on the target structures one
obtains sequence $S$=HPHPPPPHPHHHHPPH having $E_\Gamma=-13.5$. 
The true ground state configuration of $S$ has an even lower energy,
$E^\prime=-14.5$, as shown in b).}
\label{fig:test}
\end{figure}

In our lattice studies we have found considerable evidence against
this picture. We considered solutions to the design problem with
intermediate energy and selected a small number of residues. Then we
performed an energy minimization procedure over sequences that {\em
a)} had the same concentration as the reference sequence and {\em b)}
were equal to the reference sequence in correspondence of the selected
residues. In our attempts we found that very frequently the putative
solution was wrong. This is best illustrated with a simple example
given for chains of length 16 and two classes of amino acids
interacting via the NEC potentials. In Fig. \ref{fig:test}a) a ground
state conformation having energy E= -13.2 is shown. By quenching two
residues at position 4 and 14 and minimizing the energy at constant
composition one obtains the sequence HPHPPPPHPHHHHPPH. The energy of
this sequence on the original structure is E= -13.5 but its native
state is the structure shown in Fig.  \ref{fig:test}b) with energy
-14.5. This example is not an exceptional case since nearly one in 8
random attempts to quench two residues of intermediate sequences and
minimizing the energy failed to give good solutions. This is a
significant failure rate since it must be bourne in mind that nearly
one half of residues occupy ``hot'' positions that constrain them to
be of a well defined type (e.g. P residues at corners) for most
solutions. When quenching two residues at cold positions the failure
rate could be over 50\%. Quenching three key residues in chains of
length 25 may lead to a design failure rates as high as 30\%. This shows
that introducing more constraints on the designed sequence (besides
its overall composition) impairs severely the design method instead of
making it more reliable.

\section{Simple alternative design strategies}

While energy minimization methods appear to be unsatisfactory from
both theoretical and numerical points of view, they are tipically
simple and fast to implement.

On the other hand, recoursing to rigorous techniques that take into
proper account the free energy term in Eq. (\ref{eqn:occprob}) may, in
proportion, require much more CPU time since each trial solution needs
to be mounted on alternative conformations, $\{\Gamma^\prime\}$ (see
\ref{eqn:occprob}). The obvious payoff is that {\em all}\/ solutions
to design problems can be identified with $100\%$ success throughout
the whole energy range [\onlinecite{i11,i18}]. There are, however,
several design procedures based on approximate treatments of the free
energy that, while having the same speed of energy minimization
methods, are much more efficient [\onlinecite{i10,i18,M98b,S98}].

These procedures were first developed in the attempt to design real
proteins [\onlinecite{M98b}]. In that case, contrary to lattice
models, it was impossible to generate alternative off-lattice
structures competing with the target one (doing so would amount to be
able to perform a direct folding). Rather it was decided to exploit
the properties that $F(S)$ formally depends only on the sequence $S$
to expand it as a function of composition and other sequence
parameters. Contrary to energy minimization techniques this method did
not require any external intervention to fix the correct
composition. Remarkably the correct ratio of $n_H/n_P$ was
nevertheless observed in optimal solutions [\onlinecite{M98b}].

Design strategies based on a functional approach to $F(S)$ are not
only reliable, but they may even be used to determine the best
(unknown) amino acid interactions given a two-body (or higher-order)
parametrization of the Hamiltonian. Details on the use of these
methods can be found elsewhere [\onlinecite{M98b,S98}]; in the rest of
this section we will concentrate on yet another free-energy-based
design method originally proposed by Deutsch and Kurowsky
[\onlinecite{i10}]. The DK strategy is based on building a table of
the relative frequency of geometrical contacts between two residues at
sequence positions $i$ and $j$, $\langle \Delta_{ij} \rangle$,
collected over all compact structures.  Hence, for a given sequence,
$S$, the free energy can be approximated as

\begin{equation}
F_{DK}(S) = \sum_{ij} \langle \Delta_{ij} \rangle \epsilon(S_i,S_j)\ .
\end{equation}

\noindent This may be regarded as an average energy attained by
sequence $S$ on compact structures.  Requiring that $E(S) -F(S)$ is
minimized selects the sequence(s) whose energy on $\Gamma$ lies as low
as possible with respect to $F(S)$. This method can be used without
constraining the sequence composition to a particular value and its
efficiency, albeit not equal to 100 \%, is much higher than what is
obtainable by energy minimization in analogous circumstances. This was
assessed by ranking the sequences according to their normalized DK score:

\begin{equation}
{2(E(S,\Gamma)- F_{DK}(S)) \over |E(S,\Gamma)+ F_{DK}(S)|}\ ,
\end{equation}

\noindent and isolating those that stayed below a pre-assigned
threshold.  For example, for the case of length 25, NEC parameters, $
8 \le n_h \le 14$ we chose the threshold value as 0.25. The selected
sequences represented our putative solutions and were ordered
according to their ground state energy. The efficiency was calculated
as the ratio of correct solutions versus putative ones as a function
of energy. While, at low energies, the number of correct solutions was
approximately the same for the DK method and the energy minimization
one, at higher ones the efficiency of the DK procedure could be 100
times higher than the energy minimization procedure. The cumulative
efficiency of the DK approach compared to energy-minimization routines
over the whole range of energy for which
solutions existed was 20:1 (this range is easily identified with
free-energy based methods, but impossible to find out with the energy
minimization). The same proportion of efficiency was observed for
length 36, where a random sampling of sequences was performed

\section{Conclusions}

We have carried out extensive enumeration studies to assess the
performance of design techniques based on energy minimization. It is
found that these techniques can be effective in selecting correct
ground states. Nevertheless, it turns out that the overwhelming
majority of solutions do not possess minimum energy and hence cannot be
identified. It was also shown that practical implementations of
energy-based design strategies need to be more and more efficient in
finding the lowest energy solutions on increasing the sequence length
and number of amino acid classes. Finally it was found that these
methods become unreliable when additional requirements are imposed on
the properties of putative solutions, which possibly suggests that the
method may be unsuitable to design realistic proteins with desired
biological functionality.

Design techniques incorporating appropriate treatments of the free
energy do not suffer these shortcomings and can, in principle, lead to
100 \% design success at the expenses of considerable CPU time.
In the last section of this paper we discuss some approximations to
the free energy that while being as fast as energy-based methods,
appear to be more efficient. Furthermore, they do not require any prior
fixing of a correct amino acid concentration and could be used
effectively to replace energy-minimization techniques when designing
realistic proteins.

ACKNOWLEDGMENTS. We thank J. R. Banavar, R. Dima, Jort Van Mourik and
R. Zecchina for useful discussions. We are indebted to Flavio Seno for
illuminating suggestions and a careful reading of the manuscript.

\newpage

\end{document}